\documentclass[showpacs,preprintnumbers,superscriptaddress,nofootinbib,12pt]{revtex4}

\usepackage[dvips]{graphics}
\usepackage{epsfig}
\usepackage{latexsym}
\usepackage{amssymb}

\begin{document}

\title{Phenomenology of Left-Right Twin Higgs Model}


\author{Hock-Seng Goh}

\author{Shufang Su}
\affiliation{ 
Department of Physics, University of Arizona, Tucson, AZ 85721
}


\begin{abstract}
Twin Higgs mechanism has recently been proposed to solve the ``little
Hierarchy'' problem.  We studied the implementation of twin Higgs mechanism
in left-right models. We discussed the particle spectrum, and the collider
phenomenology at the Large Hadron Collider. 

\end{abstract}


\maketitle

\pagenumbering{arabic}


\section{Left-Right Twin Higgs model}

Naturalness requires the stabilization of the Higgs mass against large  
radiative corrections. The scale of new physics needs to be around the 
electroweak scale to avoid 
the fine-tuning of the Higgs potential.  On the other hand, electroweak 
precision measurements push the new physics scale to be above a few TeV.
This conflicts in the new physics energy scale  
is the so-called ``little Hierarchy'' problem.  

Recently, twin 
Higgs mechanism has been proposed as a solution to the ``little Hierarchy''
problem \cite{twinhiggsmirror, twinhiggslr}.  
The Higgs is a pseudo-Goldstone boson of a spontaneously
broken global symmetry.  
Gauge and Yukawa interactions that explicitly break  the 
global symmetry give mass to the Higgs.  
Once a discrete symmetry is imposed, the leading 
quadratic divergent term respects the global symmetry, thus does not 
contribute to the Higgs mass.  The resulting Higgs mass obtains logarithmic
corrections.  Its mass is around the electroweak scale when the 
cut off is around $5-10$ TeV.  

The twin Higgs mechanism could be implemented
in different ways.  In the mirror model \cite{twinhiggsmirror}, 
a complete copy of the 
Standard Model (SM) is introduced and 
the discrete symmetry is identified with mirror parity. 
The only collider signal for
the mirror twin Higgs model is the invisible Higgs decay, which can 
be tested at the ILC.

The twin Higgs mechanism can also be implemented in left-right models 
with the discrete symmetry being identified with left-right symmetry.
There are new particles around the electroweak scale, which interact with 
SM particles.  Such model has rich collider phenomenology, 
which will be discussed in this talk.

In the left-right twin Higgs model (LRTH), 
the global symmetry is ${\rm SU}(4)$, 
with a gauged 
${\rm SU}(2)_L\times {\rm SU}(2)_R \times {\rm U}(1)_{B-L}$ subgroup. 
The Higgs field $H=(H_L, H_R)$ is in the fundamental representation
of  ${\rm SU}(4)$, with $H_L$ charged under ${\rm SU}(2)_L$ and
$H_R$ charged under ${\rm SU}(2)_R$.  After the Higgs obtains 
a vacuum expectation
value (vev) $\langle H \rangle = (0,0,0,f_1)$, the global symmetry
${\rm SU}(4) $ breaks down to ${\rm SU}(3)$, and 
${\rm SU}(2)_R \times {\rm U}(1)_{B-L}$ breaks down to the SM ${\rm U}(1)_Y$.
Three of the seven Goldstone bosons (GB) are eaten by the 
massive gauge bosons $Z_H$ and $W_H^{\pm}$, while the rest four 
contain the SM Higgs doublet.

The fermion sector of LRTH is similar to the SM, with 
the right handed quarks $(u_R, d_R)$ and leptons $(l_R, \nu_R)$ 
charged under the ${\rm SU}(2)_R$.
Notice that we have to introduce additional right handed neutrinos.  
Their masses 
are required to be larger than the proton mass to avoid the strong 
constraints on the heavy gauge boson masses from supernovae cooling.  
To obtain order of one top Yukawa coupling, 
additional vector pair of top quark singlet, 
$T_L$ and 
$T_R$, is introduced.
The electroweak 
precision constraints push $f_1$ to be larger than 3 or 4 TeV, 
reintroducing the fine tuning of the Higgs potential. 

To fix this, another Higgs field $\hat{H}=(\hat{H}_L, 
\hat{H}_R)$ is introduced,  which {\it only}
couples to the gauge sector if we impose a matter parity.  It obtains a vev 
$\langle \hat{H} \rangle = (0,0,0,f_2)$ with  $f_2\gg f_1$.
The heavy gauge boson gets a mass of the order of $g f_2$, which 
satisfies the precision constraints.  
The fine tuning in the Higgs potential is under control since 
the gauge sector contribution is suppressed by the 
small gauge coupling, while the top sector contribution
is proportional to $f_1$, which can now be around a few hundred GeV.

The introduction of the extra Higgs field $\hat{H}$ enlarges the global
symmetry to ${\rm SU}(4)\times{\rm SU}(4)$. After spontaneous symmetry 
breaking, there are 14 GBs.  Three of the six 
GBs that are charged 
under ${\rm SU}(2)_R$ are eaten by the heavy gauge bosons, while 
leaves three physical Higgses: $\phi^0$ and $\phi^{\pm}$. 
The remaining Higgses are the SM Higgs doublet $H_L$ and 
an extra Higgs doublet $\hat{H}_L=(\hat{H}_1^+, 
\hat{H}_2^0)$ that only couples to the gauge boson sector.
A residue matter parity in the model renders the neutral Higgs 
$\hat{H}_2^0$ stable, and it could be a good dark matter  candidate.
After electroweak symmetry breaking, a physical Higgs boson 
$h_{SM}$ from $H_L$ is left, which plays the role of the SM Higgs 
particle.

\begin{figure}
\includegraphics*[width=2.5 in]{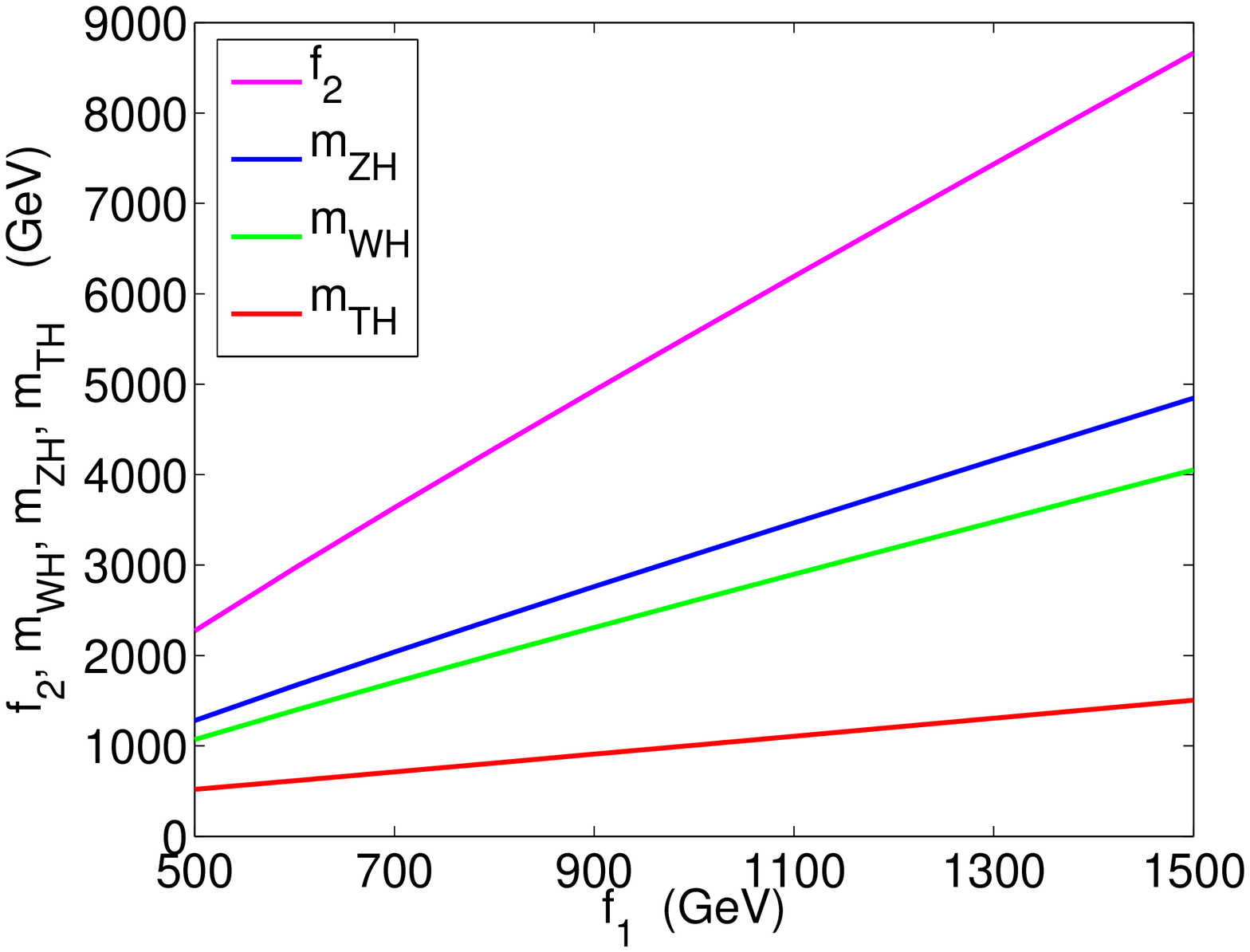}
\includegraphics*[width=2.5 in]{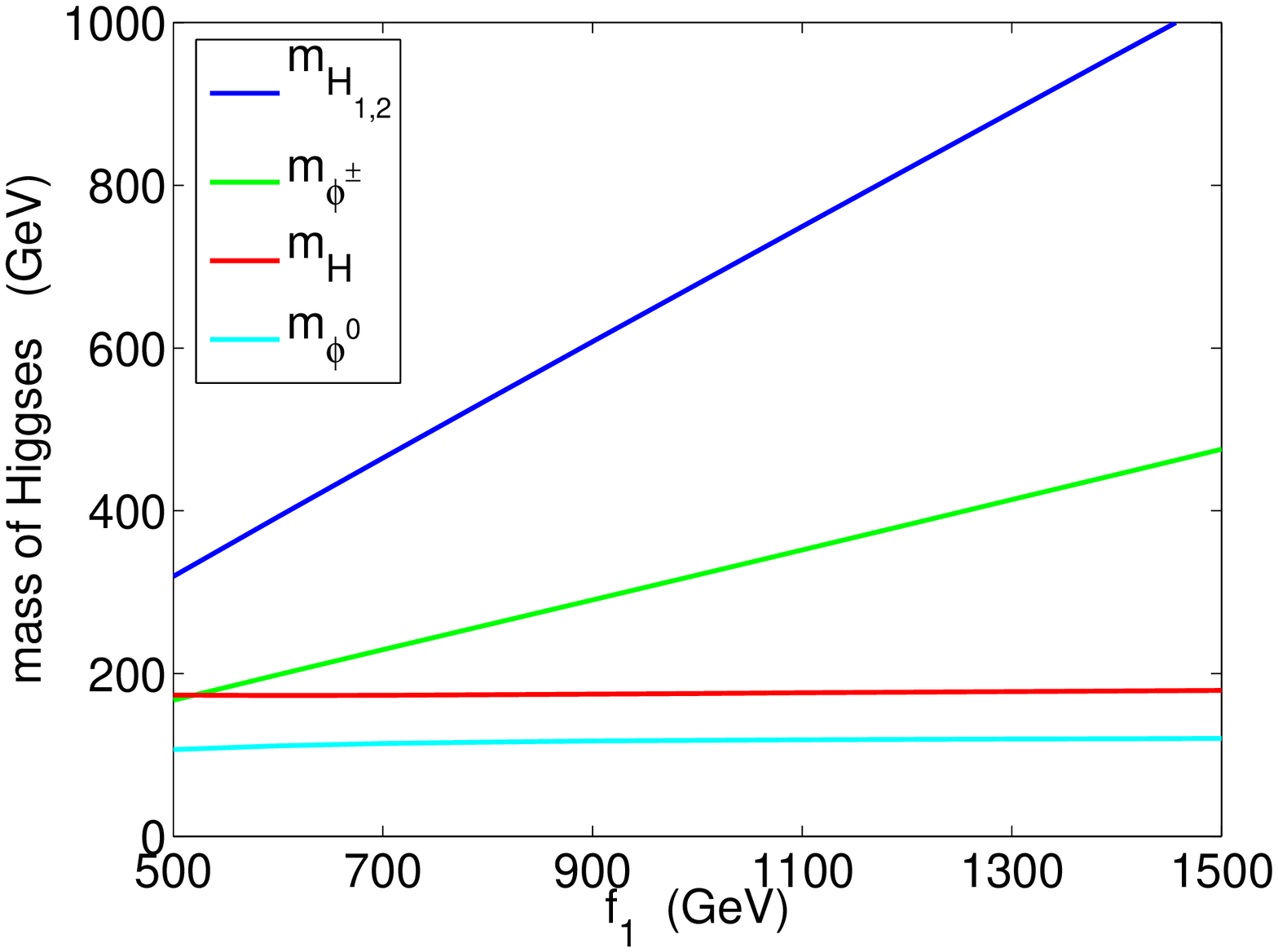}
\vspace*{-0.3 in}
\caption{The left plot shows (from top to bottom) the 
value of $f_2$ and masses of $Z_H$, $W_H$, $t_H$.  The right plot 
shows (from top to bottom) 
the masses of $\hat{H}_1^{\pm}$($\hat{H}_2^{0}$), $\phi^\pm$,
$h_{SM}$ and $\phi^0$.  The other parameters are chosen as
$\Lambda=4 \pi f_1$, $M=150$ GeV, $\sqrt{B}=50$ GeV and $\mu=f_1/2$.
}
\label{fig:mass}
\end{figure}

The new particles in the LRTH are: 
heavy gauge bosons $Z_H$, $W_H^\pm$, heavy top quark $t_H$, 
neutral Higgs $\phi^0$, a pair of charged Higgses $\phi^\pm$,
and a ${\rm SU}(2)_L$ complex Higgs doublet: $\hat{H}_1^\pm$,
$\hat{H}_2^0$.  The model parameters are 
the vevs $f_1$, $f_2$, the cut off scale 
$\Lambda$, the top quark singlet mixing parameter $M$, a mass
parameter $\sqrt{B}$ for $\phi^0$, and a  
mass parameter $\mu$ for $ \hat{H}_1^\pm$ and
$\hat{H}_2^0$.  Once $f_1$ is fixed, the vev $f_2$ can be 
determined by fixing the SM Higgs vev to be  
246 GeV.  The remaining free parameters are:
($f_1$, $\Lambda$, $M$, $\sqrt{B}$, $\mu$).

Fig.~\ref{fig:mass} shows the masses for the new particles as
a function of $f_1$, for a typical set of choice for 
($\Lambda$, $M$, $\sqrt{B}$, $\mu$).  The heavy top mass is 
between 500 GeV and 1.5 TeV.  The heavy gauge boson 
masses are above 1 TeV.  They are all within the reach of LHC. 
The mass of the Higgs $\phi^0$ is around 100 GeV for 
$\sqrt{B}=50$ GeV, while the mass of the 
charged Higgs $\phi^\pm$ is between 200 to 400 GeV. The SM Higgs mass is 
around 150 $-$ 170 GeV, the LHC reach of these particles depends on 
their production processes and decay modes, which will be discussed below.

\section{Collider Phenomenology}

\begin{figure}
\includegraphics*[width=2.5 in]{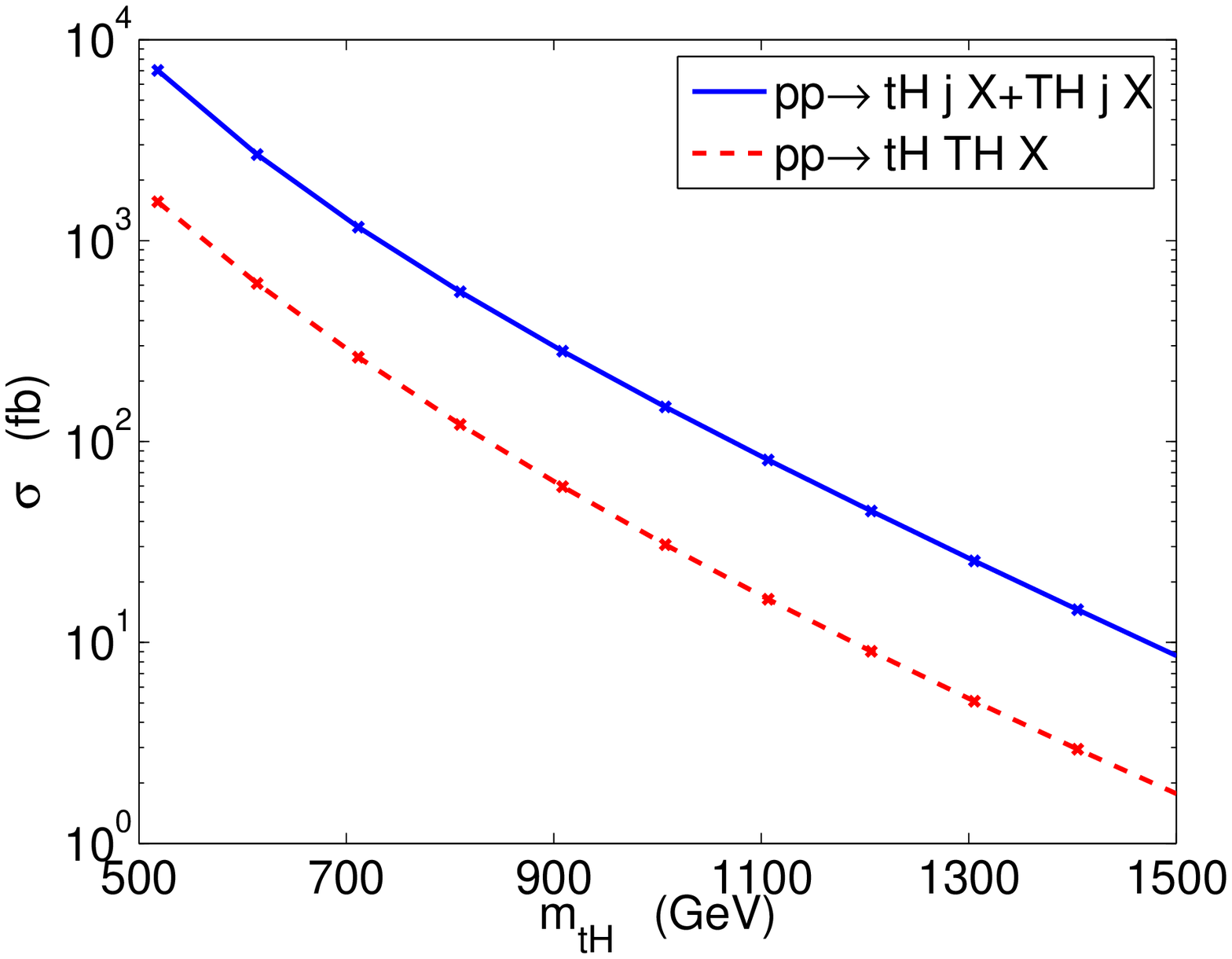}
\includegraphics*[width=2.5 in]{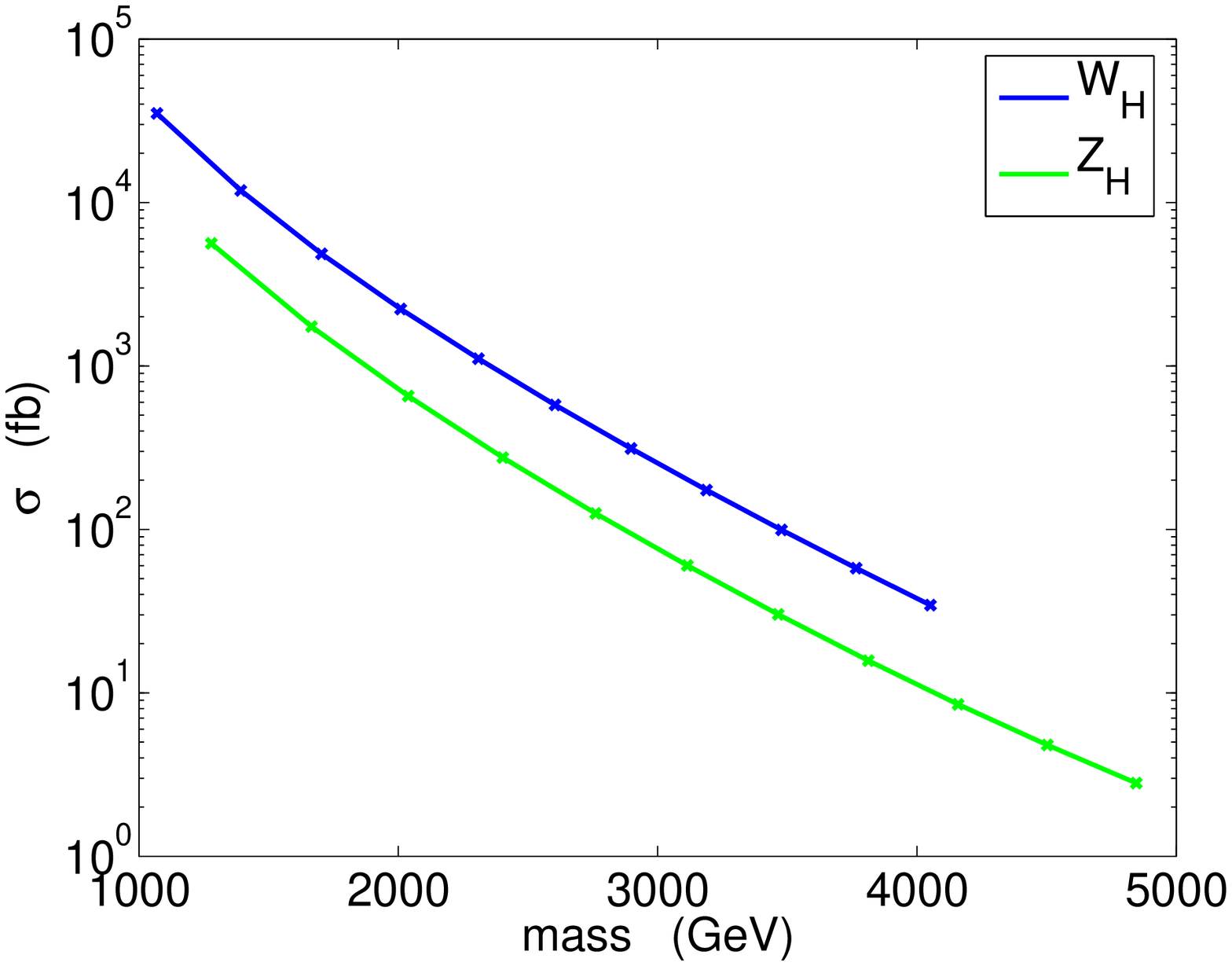}
\vspace*{-0.3 in}
\caption{Left plot shows the single and pair production for heavy top quark
at LHC.  
The right plot shows the drell-yan production cross section for $W_H$  
and $Z_H$ at LHC.
The ``x''s correspond to the 
value of $f_1$ being $100,200,...,1000$ GeV. 
The model parameters are chosen to be 
$M=150$ GeV and $\Lambda=4 \pi f_1$. 
}
\label{fig:heavytop_WHZH}
\end{figure}

The production cross section for heavy top at LHC is given 
in the left plot of Fig.~\ref{fig:heavytop_WHZH}.
The dominant production mode for heavy top at LHC is 
single heavy top production in associated with a jet: $pp\rightarrow t_H j$.
For a heavy top mass of 500 $-$ 1500 GeV, the cross section is in the 
range of $8\times 10^3$ fb $-$ 10 fb. 
The heavy top pair production via QCD process 
(dashed line in the left plot of Fig.~\ref{fig:heavytop_WHZH})
is about a factor of five smaller,
due to the phase space suppression of the large 
heavy top mass.

More than 90\% of heavy top decays via:
$t_H \rightarrow \phi^+ + b$.
Considering the subsequent decay of 
$
\phi^+ \rightarrow t \bar{b}, \ \ 
t \rightarrow W^+ b \rightarrow l^+ \nu b$,
the signal is three b-jets $+$ one charged lepton $+$ missing $E_T$.  
There is always an additional 
jet (most likely $b$-jet) that accompanies $t_H$.
A detailed collider study of this process is under current investigation.

The heavy top can also decay into $t_{SM}h_{SM}$, $t_{SM}Z$ and 
$bW$, with branching ratios of the order of $10^{-2}$ for $M=150$ GeV. 
Considering $t_H\rightarrow t_{SM}Z$ channel: 
the signal is one $b$-jet $+$ tri-lepton $+$ missing $E_T$, which is almost
SM background free.

The dominant production channels for heavy gauge bosons at hadron colliders are
the drell-yan processes: $pp\rightarrow W_H X$ and $pp\rightarrow Z_H X$, 
which are shown in the right plot of Fig.~\ref{fig:heavytop_WHZH}.
The drell-yan cross section is large: 
varying from $4 \times 10^4$ fb 
for $W_H$ mass of about 1 TeV to 40 fb for $W_H$ mass of about 4 TeV.

For $W_H$, 
the dominant decay into two jets is not very useful 
due to the huge QCD jet background. 
$W_H$ could decay into $l\nu_R$ if the right handed neutrino mass is 
less than $m_{W_H}$. It leads to a clean signal of lepton plus missing energy.
$W_H$ could also decay into $t_Hb$,
with a branching ratio of about 20\%.  Depending on the 
subsequent decays of the heavy top, 
we expect to see signals of 4 b + lepton + missing $E_T$ or 
2 b + tri-lepton + missing $E_T$.  
About 3\% $W_H$ decays into $\phi^0\phi^\pm$, which  
is the dominant production mode for $\phi^0$.

Although the dominant decay mode of $Z_H$ is into di-jets, the discovery 
modes for $Z_H$ are $t_H\bar{t}_H$ (with a branching ratio of 2-6\%), 
$t\bar{t}$ (with a branching ratio of 3-4\%) and $l^+l^-$ (with a branching 
ratio of 2.5\% for $e^+e^-$, $\mu^+\mu^-$ and $\tau^+\tau^-$ individually).  
The di-lepton mode provides a clean signal.  It 
can be separated from the SM background by studying 
the invariant di-lepton mass distribution.

The SM Higgs mass depends on $f_1$, $M$ and $\Lambda$, and is found to be 
in the  range of $150-170$ GeV. It could be discovered  via the 
gluon fusion process $gg\rightarrow H$ with Higgs decays into di-bosons.

Besides the SM Higgs, there are three additional Higgses that couple 
to both the SM fermions and gauge bosons: one neutral Higgs $\phi^0$ and
a pair of charged Higgses $\phi^\pm$.  $\phi^0$ decays into $b\bar{b}$ or
$\tau^+\tau^-$, and $\phi^\pm$ dominantly decays into $tb$.

The gluon fusion production for  $\phi^0$ is not so useful 
due to huge QCD background to the $b\bar{b}$ final states.
It could, however, be produced in the decay of heavy particles in the model.
The dominant production mode is through $W_H \rightarrow \phi^0\phi^\pm$,
with a cross section of about 1 fb $-$ $10^3$ fb.  
For $\phi^\pm$, the dominant production mode is through heavy
top decay, with a cross section in the range of 10 fb to $10^4$ fb.

The complex  Higgses $\hat{H}_1^\pm$ and $\hat{H}_2^0$ 
couple to the gauge bosons only.  Their masses are very degenerate;
$\hat{H}_1^\pm$ is slightly heavier than  $\hat{H}_2^0$ due to the 
small mass splitting introduced by the electromagnetic interactions.
$\hat{H}_1^\pm$ can therefore decay into $\hat{H}_2^0$ plus soft jets or 
leptons.  If the decay lifetime is long enough, we can see charged track
in the detector with little hadronic activity.  Otherwise, 
the soft jets and leptons can not be detected at colliders.
The events appear as missing energy, which escape the detection.

All the above discussion is for a small but non-zero $M$. 
In the limit of $M=0$, certain couplings go to zero,
which changes  the collider signatures significantly.  In particular, 
$\phi^\pm tb$ coupling is now absent, and $\phi^\pm$ can no longer
decays into $tb$ as in the non-zero $M$ case.
The only possible decay channels are $\phi^\pm \rightarrow 
\phi^0 q\bar{q}^\prime$ or
$\phi^\pm \rightarrow H q\bar{q}^\prime$.  
The former one suffers from huge QCD 
background, and the latter one suffers from small branching ratio.  
Similarly,  the discovery for all the other new particles becomes 
very challenging.  The only exceptions are $Z_H \rightarrow l^+l^-$ and 
$W_H \rightarrow l \nu_R$, which remain unaffected.

In conclusion, 
the left-right Twin Higgs model provides an alternative mechanism to solve 
the ``little Hierarchy'' problem. 
The heavy gauge bosons and heavy top partner
can be copiously produced at LHC and have rich collider phenomenology.  
The detailed collider analysis of this model is under current 
investigation.




\begin{acknowledgments}
We would like to thank Z. Chacko for useful discussions of the 
left-right Twin Higgs models.  The current work is supported by
DOE.
\end{acknowledgments}



\bibliographystyle{aipprocl} 



\end{document}